\documentclass[aip,amsmath,amssymb, reprint,cha]{revtex4-1}

\usepackage{graphicx}
\usepackage{dcolumn}
\usepackage{bm}

\usepackage[utf8]{inputenc}
\usepackage[T1]{fontenc}
\usepackage{mathptmx}

\begin{document}

\preprint{AIP/123-QED}

\title[]{Efficient moment-based approach to the simulation of infinitely many heterogeneous phase oscillators}

\author{Iv\'an Le\'on}
\author{Diego Paz\'o}
 \affiliation{Instituto de F\'isica de Cantabria (IFCA), Universidad de
 Cantabria-CSIC, Avda.~Los Castros, s/n, 39005 Santander, Spain}

\date{\today}

\begin{abstract}
The dynamics of ensembles of phase oscillators are usually described 
considering their infinite-size limit.
In practice, however, this limit is fully accessible only if the Ott-Antonsen theory can be applied,
and the heterogeneity is distributed following a rational function. 
 In this work we demonstrate the usefulness of
a moment-based scheme to reproduce the dynamics of infinitely many oscillators. 
Our analysis is particularized for Gaussian heterogeneities, 
leading to a Fourier-Hermite decomposition 
of the oscillator density. The Fourier-Hermite moments obey a set of hierarchical ordinary differential equations. 
As a preliminary experiment, the effects of truncating the moment system and implementing 
different closures are tested in the analytically solvable Kuramoto model.
The moment-based approach proves to be much more efficient than the direct 
simulation of a large oscillator ensemble.
The convenience of the moment-based approach is exploited in two illustrative examples: 
(i) the Kuramoto model 
with bimodal frequency distribution, and 
(ii) the `enlarged Kuramoto model' (endowed with nonpairwise interactions). In both systems we
obtain new results inaccessible through direct numerical integration of populations.

\end{abstract}

\maketitle

\begin{quotation}
A wide variety of systems can be modeled as ensembles of oscillators, from biological systems (e.g.~neuronal networks) to physical ones (e.g.~power grids). Their cooperative phenomena, 
e.g.~collective synchronization, 
are often described in terms of interacting (one-dimensional) phase oscillators.
The dynamics of globally coupled phase oscillators, initiated long ago by Winfree \cite{Win80} and Kuramoto \cite{Kur84},
is a vibrant topic of nonlinear science \cite{PikRos15}. 
Considering infinitely many phase oscillators (the so-called thermodynamic limit)
is a key assumption to make useful theoretical and phenomenological descriptions. 
In practice, however, simulations are customarily carried out with large ensembles
of oscillators, assuming they faithfully reflect what even larger populations do. 
This straightforward approach may be inconvenient for two reasons:
its inaccuracy due to finite-size fluctuations, and 
its computational cost due to the trigonometric functions involved.
The situation is even more pressing when collective chaos (or hyperchaos) is
found, since microscopic chaos contributes with a macroscopic amount of
positive Lyapunov exponents \cite{popovych05}, possibly masking the actual 
value of the collective exponents.
In this work, we study 
a
moment-based 
scheme for
systems of infinitely many 
heterogeneous
oscillators.
The moment system is the result of decomposing the oscillator density 
in a basis of Fourier components and orthonormal polynomials.
This decomposition appeared some years ago 
in a paper by Chiba \cite{chiba13},
although only recently it demonstrated
its utility to address numerical issues \cite{leon22}. In this work, 
we first test the moment-based approach with the Kuramoto model.
We compare different closures of the moment system. The moment-based
approach proves to be highly efficient in general. In the second part of this paper, we carry out moment-based simulations in two illustrative systems, obtaning new results hardly achievable through
the direct numerical simulation of phase oscillators. In particular,
unstable solutions can be continued uncovering in this way a previously unnoticed bifurcation,
and ``collective'' Lyapunov exponents are obtained at a relatively low computational cost.

\end{quotation}

\section{Introduction} 

Self-sustained oscillators pervade the natural world and
our technology-driven society. 
The synchronization between them (spiking cells, electronic/microwave 
circuits,  etc.) is conspicuous \cite{PRK01}. 
Ultimately, this is the consequence of natural evolution 
or technical design. Though under certain circumstances 
synchronization may also be an indicator of malfunction. 

From a mathematical perspective, the dynamics on a limit-cycle attractor 
can be parametrized by a cyclic coordinate, called the phase. 
Under weak disturbances, a perturbative technique, called phase reduction, permits 
to eliminate all degrees of freedom except the phase \cite{Kur84,nakao16}.
This automatically suggests using the phase oscillator 
as the natural unit to
describe weakly interacting limit-cycle oscillators.

Collective synchronization is typical in large ensembles of oscillators. Classical examples
include cardiac pacemaker cells in the sinoatrial node \cite{michaels87},
swarms of flashing fireflies \cite{BB76}, arrays of Josephson junctions \cite{WCS96}, to cite a few \cite{Str03}.
Populations of heterogeneous phase oscillators reproduce the
onset of collective synchronization, as originally proven in a 
seminal numerical experiment by Winfree \cite{Win67}. Some years later Kuramoto 
derived a tractable model \cite{Kur75}, subsequently used as building block
to investigate a variety of collective phenomena: 
chimera states \cite{chimera}, gamma oscillations in the brain \cite{MP18}, 
swarming \cite{swarmallators,chandra19}, etc. 
Nowadays, the dynamics of populations of phase oscillators
continue attracting the interest of a interdisciplinary
community of scientists \cite{PikRos15,BGL+20}.

The theoretical efforts to describe populations of 
globally coupled phase oscillators 
usually adopt the thermodynamic limit, i.e.~the population size $N$
is assumed to be infinite (with 
the
coupling strength scaling as $1/N$). This 
simplification is, for instance, part of the original
Kuramoto's self-consistent analysis\cite{Kur75}, 
the stability analysis of incoherence by Strogatz and Mirollo \cite{SM91},
and the Ott-Antonsen theory \cite{OA08}. 
In practice, we can generally regard the dynamics
of a large population of phase oscillators as an ideal infinite population, plus
small finite-size fluctuations vanishing in the thermodynamic limit. 
Altogether, theoretical approaches and phenomenological characterizations
routinely refer to the infinite-size limit.

In some special cases, the dynamics of infinitely many oscillators 
can be reduced to a few ordinary differential equations (ODEs). In technical terms, this
is possible when the ``Ott-Antonsen manifold'' is attracting \cite{OA08,OA09}, and
the heterogeneity is represented by a rational probability density function (e.g.~Lorentzian).
Otherwise, the standard procedure to explore the thermodynamic limit 
is simulating increasingly larger population sizes.
As we will show below this strategy is, however, not optimal and potentially inaccurate. 
It is not optimal due to 
unavoidable
finite-size fluctuations, and the computational cost of the trigonometric functions involved. 
Furthermore, within this strategy, unstable collective states are hardly accessible,
and the stability of attracting states is poorly estimated. 

The situation is even worse if the collective dynamics is chaotic. 
Direct numerical simulations may turn to be completely insufficient.  
The actual value of the Lyapunov exponents in
the thermodynamic limit may be masked
by an $O(N)$ amount of positive 
Lyapunov exponents\cite{popovych05}. We have recently found such a problem in a
population of phase oscillators with pairwise and nonpairwise interactions \cite{leon22}. 

The drawbacks of direct numerical simulations were already tackled in
the context of the Kuramoto model\cite{ABP+05} (KM).
The first step was to consider an infinite population from the 
outset, and thereby working with the oscillator density.
The most promising methodology is the moment-based approach,
initially proposed in Ref.~\cite{perezritort97}.
Here we start from a decomposition of the oscillator
density first proposed by Chiba \cite{chiba13}. In his work, the
set of moments was used as an instrument to prove that 
in the KM the limit $N\to\infty$ is well behaved.
In a recent paper\cite{leon22}, we found that the moments defined in \cite{chiba13}
serve as convenient variables for numerical simulations.
In this work, our aims
are (i) to analyze this moment-based approach in detail (assessing
the performance of different closures in the KM), and (ii)
to use moments in a couple of relevant problems (clearly outperforming
direct simulations).

The remainder of this paper is structured as follows. In Sec.~II we present 
the family of phase oscillator network models under study. 
In Sec.~III we introduce the moment-based approach, 
and derive the evolution equation of a set of Fourier-Hermite moments. 
The accuracy of the moment system in the solvable 
case of the KM is investigated in Sec.~IV. 
Sections V and VI demonstrate
the might of the moment-based approach with two examples:
the KM with bimodal frequency distribution, and the ``enlarged KM'' (a system with nonpairwise interactions).
Finally, in Sec.~VII we recapitulate our main results and 
suggest possible extensions of this work.    

\section{Globally coupled phase oscillators}

A phase oscillator is a dynamical system described solely by one cyclic variable, the phase $\theta\in[0,2\pi)$. 
As already discussed above, the study of large ensembles of globally coupled phase oscillators constitute
a popular branch of nonlinear science \cite{PikRos15}. These dynamical systems
evolve in an $N$-dimensional torus $\{\theta_j\}_{j=1,2,\ldots,N}$, and 
are particularly difficult to analyze if heterogeneity is present.  
Specifically, we consider a family of systems, whose 
deterministic evolution equations are: 
\begin{equation}\label{eqphaseosc}
	\dot{\theta}_j=\sigma\omega_j+G(\theta_j,t).
\end{equation} 
Here the overdot denotes the time derivative.
The $\omega_j$ are drawn from a probability distribution $g(\omega)$, which is centered at zero without lack of generality.
The form of Eq.~\eqref{eqphaseosc} is shared, among others, by the well-known Kuramoto-Daido system \cite{Kur84,Dai93} 
and the Winfree model \cite{Win67}.
In both situations function $G$ exclusively encodes the coupling between the oscillators, plus a frequency offset $\Omega_0$. 
Therefore $\sigma\omega_j$ are the deviations from the central natural frequency of the oscillators. In a more general setup, function $G$ may include the nonuniformity
of the rotations, as in ensembles of active rotators \cite{Sak88} (derived from a periodically forced KM).
We abuse of language and refer to $G$ as the coupling function hereafter.
In function $G$ the time dependence may enter explicitly, and implicitly
through the Kuramoto-Daido order parameters \cite{Kur75,Dai93}: $Z_k(t)=\frac{1}{N}\sum_j e^{ik\theta_j}$, $k=1,2,\ldots$.
For example, the KM 
corresponds to
$G(\theta,t)=\Omega_0+\epsilon\operatorname{Im}[Z_1(t) e^{-i\theta}]$,
where $\epsilon$ is the coupling constant. 

For convenience $g(\omega)$ is chosen to posses unit variance, 
such that parameter $\sigma$ in Eq.~\eqref{eqphaseosc} controls the dispersion. 
In this work we adopt the normal distribution $\omega_j\sim\mathcal{N}(0,1)$,
i.e.~the probability density function is:
\begin{equation}\label{gauss}
g(\omega)=\frac1{\sqrt{2\pi}} e^{-\omega^2/2} . 
\end{equation}
Other distributions with finite moments can, in principle, be analyzed in an analogous way,
as discussed in Sec.~VII.

\section{Theory}\label{sectModes}

\subsection{Continuous formulation}

The dynamics of large systems of the form \eqref{eqphaseosc} is usually described assuming 
the thermodynamic limit, $N\to\infty$. It is seldom proven \cite{chiba13}, but
usually assumed, that 
for sufficiently large system sizes the dynamics is simply that 
of the thermodynamic limit, supplemented by irrelevant, asymptotically small, finite-size fluctuations
\footnote{Throughout this work the equivalence between continuous 
and finite-$N$ formulations is taken for granted (as proven, for example, 
for the KM in \cite{chiba13}). This is not always the case though. In a model
analyzed in \cite{KP15}, ${\dot \theta}_j=\sigma\omega_j+\epsilon R^2
\sin(2\Psi-2\theta_j)$, incoherence remains unstable for any finite $N$
(and large enough $\epsilon$), while it is stable in the continuum limit. 
For finite populations residence times near incoherence diverge with $N$. 
Such, perhaps pathological, situations require a careful analysis and are ignored hereafter.}. 
Still, the analysis of the thermodynamic limit of Eq.~\eqref{eqphaseosc} is not trivial in general. 
The only analytically solvable situation is that in which $G$ contains only the first harmonic in $\theta$
(often expressed as $G(\theta,t)=\operatorname{Im}[H(t) e^{-i\theta}]$),
and $g(\omega)$ is a rational function.  
The dynamics becomes exactly described by a few ODEs
inside the Ott-Antonsen manifold\cite{OA08} (but not the transient
from an arbitrary initial condition).

Simulating Eq.~\eqref{eqphaseosc} for an increasing number of oscillators may give an idea of the
asymptotic dynamics in the thermodynamic limit.  
Alternatively, we may choose to start the analysis in the thermodynamic limit, 
defining a conditional oscillator density $\rho(\theta|\omega,t)$ 
such that $\rho(\theta|\omega,t)g(\omega)d\theta d\omega$ is the fraction of oscillators 
with phases between $\theta$ and $\theta+d\theta$ and
``pseudo-frequencies'' between $\omega$ and $\omega+d\omega$ at time $t$. The density $\rho$ obeys the continuity equation:
\begin{equation}\label{cont}
 \frac{\partial\rho}{\partial t}=-\frac{\partial}{\partial\theta}\left\{\left[\sigma\omega+G(\theta,t)\right] \rho  \right\} .
\end{equation}
A variety of approaches have been proposed in order to efficiently solve the previous equation, specially in the context
of the KM, see Sec.~VI.B in \cite{ABP+05}.

\subsection{P\'erez-Vicente and Ritort's moments}

The moment-based approach was inaugurated in 1997 by P\'erez-Vicente and Ritort 
\cite{perezritort97}. They
put forward a set of modes $H_k^m$, where $k$ and $m$ are integers ($m\ge0$). In terms
of the oscillator density
\begin{equation}\label{oldmodes}
H^m_k(t)=
\int_0^{2\pi} d\theta e^{ik\theta} \int_{-\infty}^\infty  d\omega 
 g(\omega) \omega^m \rho(\theta|\omega,t)  .
\end{equation}
The hierarchical set of ODEs governing the moments $H_k^m$ 
can be obtained from Eq.~\eqref{cont}.
This numerical scheme was successfully tested in \cite{perezritort97} with the Kuramoto
model with a bi-delta distribution of natural frequencies (and white noise, which is trivially
incorporated, see Sec.~VII). 
Its performance with a continuous frequency distribution remains, to our knowledge, unknown.

\subsection{Chiba's moments: Orthonormal polynomials}

Another moment system was introduced by Chiba \cite{chiba13} in 2013,
with the aim of proving that the dynamics of the $N$-dimensional Kuramoto
model converges to the continuous model as $N\to\infty$.
In Ref.~\cite{chiba13} the moments were defined as
\begin{equation}\label{defmod}
P^m_k(t)=
\int_0^{2\pi} d\theta e^{ik\theta} \int_{-\infty}^\infty  d\omega 
g(\omega) h_m(\omega)  \rho(\theta|\omega,t)  .
\end{equation}
Instead of powers of $\omega$, as in \eqref{oldmodes}, now the
definition in \eqref{defmod} includes a function $h_m(\omega)$, which is one element of 
the orthonormal set of polynomials satisfying
\begin{equation}
\int_{-\infty}^\infty h_m(\omega) h_n(\omega) g(\omega) d\omega=\delta_{m,n} .
\end{equation}
For Gaussian $g(\omega)$, the appropriate basis is formed by
the probabilist's Hermite polynomials (rescaled by $\sqrt{m!}$): $h_m(x)=\mathrm{He}_m(x)/\sqrt{m!}$.
The conditional oscillator density is expanded in the basis of the Fourier-Hermite modes \eqref{defmod} as
follows:
	\begin{equation}\label{expansion}
	 \rho(\theta|\omega,t)=
	 \frac{1}{2\pi} \sum_{k=-\infty}^\infty \sum_{m=0}^\infty 
	P_k^m(t) e^{-ik\theta} h_m(\omega) .
	\end{equation}
	
The Fourier-Hermite modes $P_k^m$ appear to be particularly convenient. 
They are the extension of the Kuramoto-Daido order parameters to the space of the
natural frequencies. In particular, $P_k^0=Z_k$ in the thermodynamic limit. 
Moreover, $P_0^m=\delta_{m,0}$, in contrast to the analogous modes $H_0^m$ in \eqref{oldmodes}, which are nonzero.
This implies that in a uniform incoherent state (UIS), $\rho=1/(2\pi)$, the only nonzero moment is $P_0^0=1$.

\subsection{Evolution equation of the Fourier-Hermite modes}

The  ODEs governing the dynamics of the $P_k^m$ modes 
are obtained inserting \eqref{expansion}  
into the continuity equation \eqref{cont}.
As a preliminary step, we write the Fourier decomposition of the coupling function $G(\theta,t)$:
 \begin{equation} \label{gl}
 G(\theta,t)=\frac1{2\pi} \sum_{l=-\infty}^\infty G_l(t) e^{-il\theta}
 \end{equation}
where $G_l=G_{-l}^*$, as $G$ is a real-valued function. 
The right-hand side of \eqref{cont} yields two terms. The first one is:
\begin{equation}
\frac{\partial(\sigma \omega \rho)}{\partial\theta}=-\frac{i\sigma}{2\pi}\sum_{k,m} k e^{-ik\theta} P_k^m [\sqrt{m} h_{m-1}+\sqrt{m+1} h_{m+1}] ,
\end{equation}
where we have used the recurrence relation $\omega h_m(\omega)=\sqrt{m} h_{m-1}(\omega)+\sqrt{m+1} h_{m+1}(\omega)$. The second term in \eqref{cont} is:
\begin{equation}
\frac{\partial(G \rho)}{\partial\theta}= \frac{-i}{2\pi} \sum_{k,m,l} (k+l) e^{-i(k+l)\theta} G_l P_k^m  h_m .
\end{equation}
Collecting terms accompanying $e^{-ik\theta}h_m(\omega)$ at both sides of Eq.~\eqref{cont}, we get:
\begin{multline}
\frac{1}{k}\dot{P}_k^m= i\sigma\left(\sqrt{m} P_k^{m-1}+\sqrt{m+1}P_k^{m+1}\right)+
i \sum_{l=-\infty}^{\infty} P_{k-l}^m G_l,
\end{multline}
where it is implicit that $P_{-k}^m=(P_k^m)^*$. For clarity, we can 
split the last sum in the previous equation: 
\begin{multline}
\frac{1}{k}\dot{P_k^m}= i\sigma\left(\sqrt{m} P_k^{m-1}+\sqrt{m+1}P_k^{m+1}\right)\\
+iP_k^mG_0+\sum_{l=1}^{\infty}iP_{k+l}^m G_{l}^*+iP_{k-l}^mG_l .
\end{multline}
This equation becomes slightly simplified 
rotating each moment by $-\pi m /2$ radians: 
\begin{equation}
\mathsf{P}_k^m=(-i)^mP_k^m .
\end{equation}
Then, the resulting system of complex-valued ODEs reads
\begin{eqnarray}\label{eqdotmodes}
\frac{1}{k}\dot{\mathsf{P}_k^m}&=& \sigma\left(\sqrt{m} \mathsf{P}_k^{m-1}-\sqrt{m+1}\mathsf{P}_k^{m+1}\right) \nonumber\\
&+&i\mathsf{P}_k^mG_0+\sum_{n=1}^{\infty}i\mathsf{P}_{k+n}^m G_{n}^*+i\mathsf{P}_{k-n}^mG_n .
\end{eqnarray}
This infinite  set of ODEs 
exactly describes the dynamics of \eqref{cont}. 
In the limit $\sigma\rightarrow 0$, the modes decouple in the index $m$, and the dynamics is fully described
by the subset $\mathsf{P}_k^0=Z_k$, i.e.~the usual Fourier modes for a homogeneous system \cite{ShiKur86}.

\subsection{Closure}

Equation \eqref{eqdotmodes} is useful as long as working with a finite number of moments in the numerical simulation
is able to reproduce the dynamics of the oscillator ensemble, Eq.~\eqref{eqphaseosc}. 
The expectation is that (with a suitable truncation) Eq.~\eqref{eqdotmodes} should capture the thermodynamic limit 
of \eqref{eqphaseosc} better than a direct simulation of a large number of oscillators. In the latter case 
finite-size fluctuations systematically deteriorate the results.
 
\begin{figure*}
	\includegraphics[width=\linewidth]{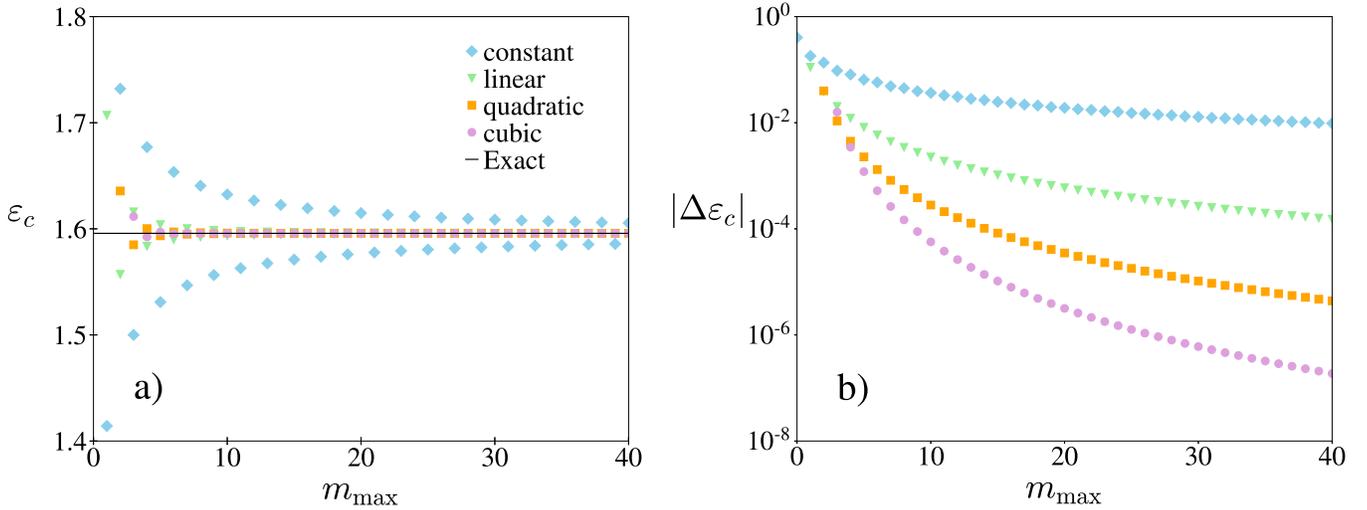}
	\caption{(a) Instability threshold of incoherence $\epsilon_c$ for
	the moment system \eqref{eqmodeskur} as a function of $m_{\mathrm{max}}$ ($\sigma=1$).
	Four different polynomial closures are considered: $n_a=0$, constant; $n_a=1$, linear; etc. 
	The exact value of $\epsilon_c$ in the thermodynamic limit is marked by a horizontal solid line.
	(b) Log-linear plot of the deviation from 
	the exact result for the data sets in panel (a).}
	\label{figuis}
\end{figure*}

We restrict here to rectangular truncations of the modes, i.e.~we neglect all the modes with 
$k>k_\mathrm{max}$ and $m>m_\mathrm{max}$. Therefore, we are confronted with dynamical systems of dimension
$2\times k_\mathrm{max}\times (m_\mathrm{max}+1)$.
In practical terms, the values of $k_\mathrm{max}$ and $m_\mathrm{max}$ are selected as a 
trade-off between the numerical accuracy 
and the computational capabilities. For the problems treated in this paper, 
the terms $G_l$ in the Fourier expansion of $G$ beyond $|l|=1$ or $2$ vanish. 
Then, values of $k_\mathrm{max}$ and $m_\mathrm{max}$
of the order of a few tens are typically enough, while keeping reasonable running times in a desktop computer.
Physical considerations may require to increase $k_\mathrm{max}$ or $m_\mathrm{max}$. For instance, 
if the system exhibits a very coherent state $|Z_1|\approx 1$ the value of $k_\mathrm{max}$ may need to be larger.
In addition, $m_\mathrm{max}$ may need to be increased if the dynamics becomes particularly complex.

The second point to be addressed is the boundary condition imposed when truncating Eq.~\eqref{eqdotmodes}.
Concerning the index $k$, we simply impose $P_{k_\mathrm{max}+1}^m=0$, in analogy to the condition
used for homogeneous systems \cite{ShiKur86}. For the index $m$ choosing a
suitable closure is not obvious. As shown below, 
imposing $P_k^{m_\mathrm{max}+1}=0$ yields inconsistent 
results. We are therefore impelled to refine the closure. 
 A constant boundary condition $\mathsf{P}_k^{m_{max}+1}=\mathsf{P}_k^{m_{max}}$,
or a linear extrapolation 
\begin{equation}\label{linear}
\mathsf{P}_k^{m_{max}+1}=2\mathsf{P}_k^{m_{max}}-\mathsf{P}_k^{m_{max}-1}
\end{equation}
are obvious candidates.
 In general, we can take an
 extrapolation of $\mathsf{P}_k^{m_{max}+1}$ using
a polynomial function of degree $n_a$. The general formula is
\begin{equation} \label{closures}
 \mathsf{P}_k^{m_\mathrm{max}+1}=\sum_{n=1}^{n_a+1}\binom{n_a+1}{n}(-1)^{(n+1)}\mathsf{P}_k^{m_\mathrm{max}-n+1},
 \end{equation}
 where $n_a=0$, 1, 2, and 3 correspond to constant, linear, quadratic, and cubic degrees, respectively.


\section{Kuramoto model with Gaussian frequency distribution}

The KM with unimodal frequency distribution 
is a paradigmatic  example of globally coupled oscillator system, 
which is analytically solvable to a large extent. We resort to it as 
a preliminary testbed system, where Eq.~\eqref{eqdotmodes} and their closures in Eq.~\eqref{closures}
can be examined in detail.

As already advanced, in the KM the interaction function is:
\begin{equation}\label{kuramoto}
G(\theta,t)=\epsilon R \sin(\Psi-\theta)=\frac{\epsilon }{2i} (Z_1 e^{-i\theta}-Z_1^*e^{i\theta}) ,
\end{equation}
where $Z_1\equiv R \, e^{i\psi}$ is the Kuramoto order parameter, and the central frequency $\Omega_0$ 
was 
set equal to
zero by going to a rotating frame. 
At low couplings, $0<\epsilon<\epsilon_c$, the KM exhibits the uniform incoherent state (UIS),
characterized by a uniform oscillator density. In the UIS all  Kuramoto-Daido 
order parameters vanish,  $Z_k=0$ (in the thermodynamic limit, otherwise fluctuations around zero survive). 
Above the critical coupling, $\epsilon>\epsilon_c$, 
a state of partial synchrony (PS) spontaneously sets in. In the state of PS 
a macroscopic cluster of oscillators becomes phase-locked, and $Z_k\ne0$ accordingly.

With the interaction function in Eq.~\eqref{kuramoto}, Eq.~\eqref{eqdotmodes}
governing the evolution of the Fourier-Hermite modes becomes:

	\begin{multline}\label{eqmodeskur}
	\frac{1}{k}\dot{\mathsf{P}}_k^m= \sigma\left(\sqrt{m} \mathsf{P}_k^{m-1}-\sqrt{m+1}\mathsf{P}_k^{m+1}\right)\\
	+\frac{\epsilon}{2}\bigg(\mathsf{P}_{k+1}^mZ_1^*-\mathsf{P}_{k-1}^mZ_1\bigg) .
	\end{multline}

\subsection{Critical coupling}

As a first test, we check if Eq.~\eqref{eqmodeskur} with different closures 
is able to reproduce the instability of incoherence at the critical coupling $\epsilon_c$.
With Gaussian $g(\omega)$, Eq.~\eqref{gauss}, the critical coupling satisfies 
$\epsilon_c/\sigma=\sqrt{8/\pi}=1.595769\ldots$.

We linearize Eq.~\eqref{eqdotmodes} around the UIS,
corresponding to $\mathsf{P}_k^m=\delta_{k,0}\delta_{m,0}$. Only the modes with $k=1$ may destabilize. 
The corresponding (linearized) ODEs read:
	\begin{equation}\label{eqmodeskurlin}
	{\dot{\mathsf{P}}_1^m}= \sigma\left(\sqrt{m} \mathsf{P}_1^{m-1}-\sqrt{m+1}\mathsf{P}_1^{m+1}\right)
	-\frac{\epsilon}{2} \delta_{m,0} \mathsf{P}_1^0 .
	\end{equation}
The critical coupling $\epsilon_c$ at which one eigenvalue of the Jacobian matrix crosses zero
depends on $m_{\mathrm{max}}$ and on the boundary condition. The null ``boundary condition'' $P_1^{m_{\mathrm{max}}+1}=0$
turns out to be inadequate, since then the critical coupling is predicted to be zero, irrespective of the value of $m_{\mathrm{max}}$. 
For other choices, see Eq.~\eqref{closures}, the value of $\epsilon_c$ converges to the
exact result as $m_{\mathrm{max}}$ grows. In Fig.~\ref{figuis}(a) we may see that the value
of $\epsilon_c(m_{\mathrm{max}})$ converges to the asymptotic value in a non-monotonic way.
The larger the degree of the polynomial extrapolation ($n_a$), the faster the convergence, see 
Fig.~\ref{figuis}(b). Empirically, we find the 
power-law convergence $|\Delta\epsilon_c|\equiv|\epsilon_c-\epsilon_c(m_{\mathrm{max}})|\propto (m_\mathrm{max})^{-n_a-1}$.  
In the $N$-dimensional KM The convergence is comparatively 
slower with the number of degrees of freedom:
the onset of entrainment spreads over a distance
$\delta\epsilon_c\sim N^{-2/5}$ for randomly sampled natural frequency distribution \cite{hong07},
while $\delta\epsilon_c\sim N^{-4/5}$ for deterministic samplings \cite{hong15}.

\subsection{Partially synchronized state}

The analytical tractability of the KM 
permits to obtain the exact values of the Fourier-Hermite modes in the PS state.
We use them as a reference for comparing the accuracy of
different truncations and closures of the moment system.
We restrict our study to a specific coupling above criticality $\epsilon/\sigma=1.8$.
The stationary density of PS can be written as a Fourier 
expansion
\begin{equation}
\rho(\theta|\omega)=\frac{1}{2\pi}+\frac{1}{2\pi}\left[\sum_{k=1}^\infty \alpha(\omega)^k e^{-ik\theta}  + \mathrm{c.c.}\right],
\end{equation}
where c.c.~stands for complex conjugate. The coefficients are $\alpha(\omega)^k$, as noted by Ott and Antonsen
when their ansatz was proposed \cite{OA08}. Function $\alpha$ 
is piecewise defined distinguishing between oscillators locked to the mean field
and drifting oscillators ($\sigma=1$ is adopted hereafter):
\begin{equation}
\alpha(\omega)=
\left\{
\begin{array}{ccc}
\sqrt{1-\frac{\omega^2}{\epsilon^2R^2}}+i\frac{\omega}{\epsilon R}& \text{if} & |\omega|\leq \epsilon R\\
\frac{i \omega}{\epsilon R}\left(1-\sqrt{1-\frac{\epsilon^2 R^2}{\omega^2}}\right)& \text{if}  
& |\omega|\ge \epsilon R
\end{array}\right.
\end{equation}
where we have chosen a reference frame such that $Z_1=R>0$.
As a preliminary step, we determined $R$ solving the self-consistence 
condition
\footnote{This condition can be expressed in terms of the modified Bessel functions
$1=\epsilon \sqrt{\frac{\pi }{8}} e^{-R^2 \epsilon ^2/4} 
\left[I_0\left(\frac{R^2 \epsilon ^2}{4}\right)+I_1\left(\frac{R^2 \epsilon ^2}{4}\right)\right]$.}
$R=\int_{-\infty}^\infty  d\omega 
g(\omega)  \alpha(\omega)$.
Now the Fourier-Hermite modes are obtained from Eq.~\eqref{defmod} integrating the powers of $\alpha(\omega)$
\begin{equation}\label{pkmexact}
P^m_k(t)=
\int_{-\infty}^\infty  d\omega \,
g(\omega) \, h_m(\omega) \,  \alpha(\omega)^k  .
\end{equation}

\begin{figure*}
	\includegraphics[width=\linewidth]{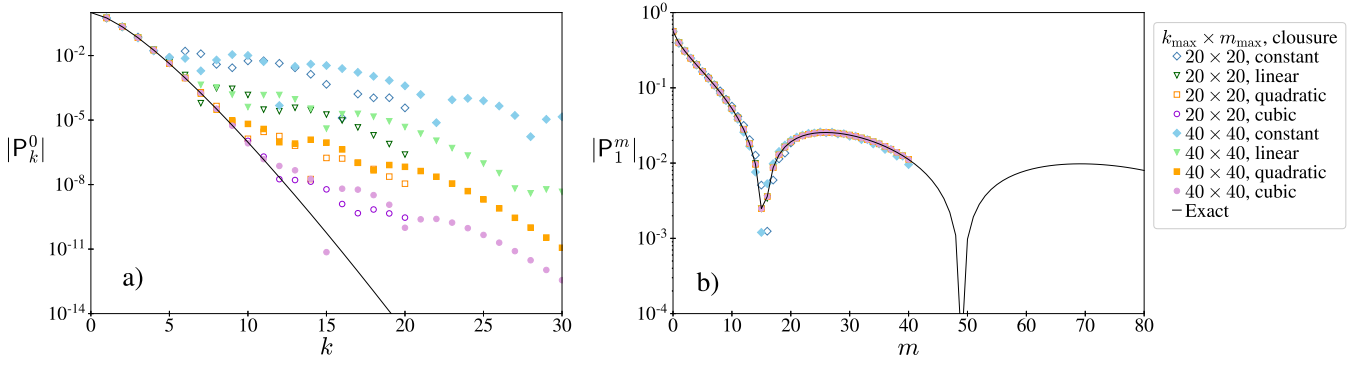}
	\caption{Absolute value of the Fourier-Hermite modes $|\mathsf{P}_k^m|$ in the PS state for $\epsilon=1.8$. 
	The exact results are depicted as a solid line. Symbols correspond to modes obtained from a truncation of Eq.~\eqref{eqmodeskur} 
	with the configurations indicated in the label at the right-hand side. (a) Constant $m=0$, (b) Constant $k=1$.
	}
	\label{figmodes}
\end{figure*}

The numerical evaluation of the integral in Eq.~\eqref{pkmexact} yields the exact values of the Fourier-Hermite 
mode $P_k^m$ (we did not cast the solution in terms of non-elementary functions).
For the specific coupling constant $\epsilon=1.8$, the absolute values of the modes with $m=0$
are depicted in Fig.~\ref{figmodes}(a) by a solid black line.
The magnitudes of the modes eventually decrease at an exponential rate with $k$. 
For other values of $m$ (not shown), the exponential regime is reached after an initial growth at small $k$ values.
The dependence of the modes on the index $m$ is more convoluted.
Figure \ref{figmodes}(b) shows, as a solid line, the representative case $k=1$. The decay with index $m$ exhibits
damped oscillations, and an envelope decaying as a stretched exponential $e^{-b\sqrt{m}}$ according to our numerical
exploration.

Figure \ref{figmodes} also shows the Fourier-Hermite modes obtained as the fixed point
of \eqref{eqmodeskur} with four different closures, and two different system sizes ($k_\mathrm{max}=m_\mathrm{max}=20$ or $40$.) 
A Newton-Raphson algorithm was used to locate the fixed point in each case. (The result agrees with the
attracting resting state observed simulating Eq.~\eqref{eqmodeskur}.)
As occurred with the critical coupling, the larger 
the degree of the polynomial extrapolation, 
the more accurate the results are. 
In Fig.~\ref{figmodes}(b) only the constant boundary condition appreciably deviates from the exact result.
The linear extrapolation \eqref{linear}
is used in the rest of the paper as it represents a good tradeoff between accuracy and simplicity. The
quadratic and cubic boundary conditions have proven to be more accurate so far.
However, this advantage ---proven for time-independent quantities---
may not 
materialize
when (complex) dynamics arise,
due to the risk of overfitting.
	
\begin{figure}
	\includegraphics[width=\linewidth]{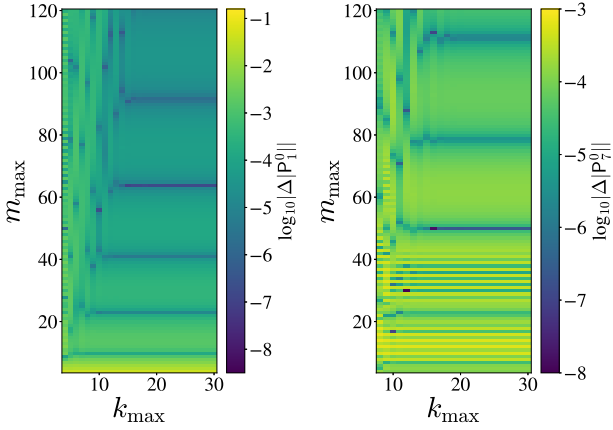}
	\caption{Level plot of the error of $|\mathsf{P}_1^0|$ (left) and $|\mathsf{P}_7^0|$ (right)
	as a function of $k_\mathrm{max}$ and $m_\mathrm{max}$, for the linear closure \eqref{linear}. 
	The scale is logarithmic, see color bar adjacent to each panel. 
	The KM is in the state of PS with $\epsilon=1.8>\epsilon_c$. The exact values of 
	the modes are $|\mathsf{P}_1^0|=0.563\ldots$ and $|\mathsf{P}_7^0|=1.79\ldots\times10^{-4}$.
	}
	\label{figerror}
\end{figure}

In Fig.~\ref{figmodes}, the number of modes ($20\times20$ or $40\times40$) appears not 
to be very relevant. This somewhat unexpected conclusion deserves to be
analyzed systematically.
We conclude this section with an extensive exploration 
of the impact of $k_\mathrm{max}$ and $m_\mathrm{max}$.
In Fig.~\ref{figerror}, the error of the modulus of the Kuramoto-Daido order parameters $|Z_1|=|\mathsf{P}_1^0|$ and $|Z_7|=|\mathsf{P}_7^0|$, 
for different configurations of $k_\mathrm{max}$ and
$m_\mathrm{max}$, are depicted in color scale. The figure shows that a minimum number of modes in $k$ and $m$ are needed to obtain a reliable result. 
With a few hundred modes ($m_{max}>k_{max}\sim20$) the errors already become quite small. The figure
clearly confirms that the error eventually decreases upon increasing $k_\mathrm{max}$ and $m_\mathrm{max}$
(though in a specific, non-monotonic fashion for each mode). 

\section{Kuramoto model with bimodal distribution}

The main goal of this section, and the next one, is to illustrate the practicality of 
the Fourier-Hermite modes with a couple of systems poorly described through a brute-force 
numerical integration of the oscillator ensembles.

The KM with bi-modal frequency distribution is a classical problem \cite{ABP+05,Str00}.
With the advent of the Ott-Antonsen theory, the problem became (almost) fully
solvable for frequency distributions of rational type, see e.g.~\cite{MBS+09,PM09,pietras18,guo20}.
Instead, we consider here a distribution equal to the sum of two normal 
distributions centered at $\pm\Omega_0$ and variance $\sigma^2$. 
Individual natural frequencies $\Omega_j$ are  distributed as:
$\Omega_j\sim[\mathcal{N}(\Omega_0,\sigma^2)+\mathcal{N}(-\Omega_0,\sigma^2)]/2$.
This distribution, bimodal only if $\Omega_0>\sigma$, 
was previously studied in Ref.~\cite{MBS+09}. There, several domain boundaries in the phase diagram 
were determined imprecisely, since they were obtained from simulations with 
$N=10000$ 
oscillators. 
This problem is a good example to demonstrate the potential of the Fourier-Hermite moments.

For convenience, we reformulate the system as a two-population problem,
i.e.~as two populations with Gaussian frequency distributions centered at $\pm\Omega_0$
with variance $\sigma^2$.
Denoting the phases of each subpopulation as $\theta^+_j$ and $\theta^-_j$,
we can write the ODEs governing the system as:
	\begin{equation} \label{eqphasebimodal}
	\dot{\theta}^{\pm}_j=\sigma\omega_j\pm\Omega_0+\epsilon R\sin(\Psi-\theta^\pm_j) ,
	\end{equation}
where $Z_1\equiv R \, e^{i\Psi}$ is the average over the order parameters
of each subpopulation: $Z_1=(Z_1^+ +Z_1^-)/2$.
We introduce two sets of Fourier-Hermite modes, $\mathsf{P}_k^m$ and $\mathsf{Q}_k^m$,
for the subpopulations centered at $\Omega_0$ and $-\Omega_0$, respectively. 
After a straightforward calculation, cf.~\eqref{eqmodeskur}, we obtain the evolution equation of 
each set:
		\begin{subequations}\label{eqmodesbimodal}
\begin{eqnarray}
			\frac{1}{k}\dot{\mathsf{P}}_k^{m}&=&i\Omega_0\mathsf{P}_k^m+f(Z_1,P_{k\pm 1}^{m\pm 1}) , \\
			\frac{1}{k}\dot{\mathsf{Q}}_k^{m}&=&-i\Omega_0\mathsf{Q}_k^m+f(Z_1,\mathsf{Q}_{k\pm 1}^{m\pm 1}),
\end{eqnarray}
		\end{subequations}
where the function $f$ is a shorthand notation for
	\begin{eqnarray}
	f(Z_1,\mathsf{P}_{k\pm 1}^{m\pm 1})=\sigma\left(\sqrt{m} \mathsf{P}_k^{m-1}-\sqrt{m+1}\mathsf{P}_k^{m+1}\right) \nonumber\\
	+\frac{\epsilon}{2}\bigg(\mathsf{P}_{k+1}^{m}Z_1^*-\mathsf{P}_{k-1}^{m}Z_1\bigg) .
	\end{eqnarray}
The identity $Z_1=(\mathsf{P}_1^0+\mathsf{Q}_1^0)/2$ closes the system of equations.

\begin{figure*}
	\includegraphics[width=\linewidth]{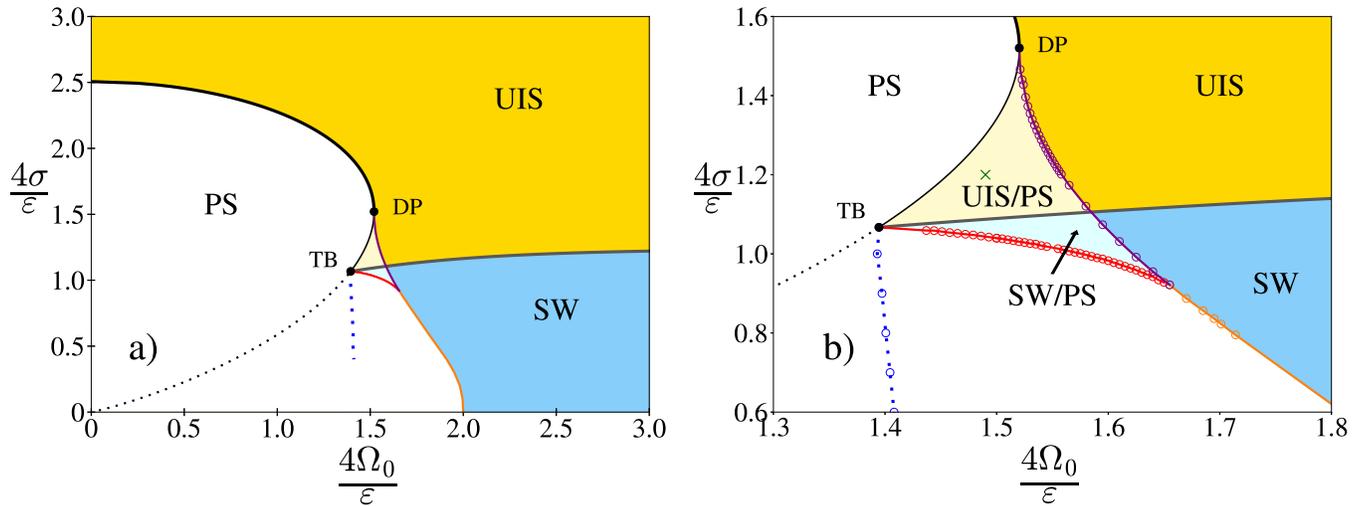}
	\caption{(a) Phase diagram of the KM with bi-Gaussian frequency distribution. 
	The domains of UIS and SW are shaded in yellow and cyan, respectively. Light shading is used to signify
	coexistence with PS.
	The black lines indicate the locus of the circle-pitchfork bifurcation of incoherence. 
	The solid style is used when an attractor is involved in the bifurcation; 
	the thick (thin) line correspond to the super- (sub-) critical bifurcation of UIS. 
	The gray line is the locus of a (degenerate) Hopf bifurcation, UIS$\rightarrow$SW transition. Both circle-pitchfork and Hopf lines are computed analytically. The pitchfork bifurcation line has a closed
	expression $\tilde\Omega_0^2=-2\tilde \sigma_P^2\ln(\tilde\sigma_P/\sqrt{2\pi})$, with $\tilde\sigma=4\sigma/\epsilon$, and $\tilde\Omega_0=4\Omega_0/\epsilon$.
		Codimension-2 points DP (degenerate pitchfork), and TB (Takens-Bogdanov) fall on top of this line
		at $\tilde\sigma=\tilde\Omega_0=1.52035\ldots$, and at
        $\tilde\sigma=1.06706\ldots$, respectively.
	The purple, red, and orange solid lines correspond to numerical estimations, via Eq.~\eqref{eqmodesbimodal}, 
	of saddle-node, SNIC and homoclinic bifurcations, respectively. 	
	A third codimension-2 point is found at the coalescence of
	the saddle-node and the homoclinic lines.
	Dotted lines correspond to bifurcations involving unstable states only. In particular, the blue dotted line
	is the (partial) locus of a drift-pitchfork bifurcation at which mirror
	traveling waves annihilate.
	(b) Magnification of the bistability regions in panel (a). Numerical data are depicted as empty circles. 
	The cross in the UIS/PS region indicates the parameter values used in Fig.~\ref{figfluct}.
	}
	\label{figbimodal}
\end{figure*}

In Fig.~\ref{figbimodal} we show the phase diagram of the model. For ease of comparison, the scaling of the axes is 
identical to Fig.~6 in \cite{MBS+09}. The model exhibits three different behaviors:
the uniform incoherent state (UIS), partial synchronization (PS), and standing wave (SW). 
The latter state corresponds to two counter-rotating clusters
of phase-locked oscillators. Moreover, two adjacent regions of bistability exist: UIS/PS and SW/PS.
Solid lines in black and grey constitute the stability
boundary of UIS, and they correspond to (circle-)pitchfork, and (degenerate) Hopf bifurcations, respectively.
These lines are obtained analytically, and are therefore identical to those in Ref.~\cite{MBS+09}. 

In contrast, the colored solid lines in Fig.~\ref{figbimodal} 
can only be obtained numerically. They correspond to standard saddle-node (purple), saddle-node on the invariant circle
(SNIC, orange), and homclinic (red) bifurcations. The location of these lines is remarkably improved
with respect to Ref.~\cite{MBS+09}.
Our numerical simulations were carried out with $k_\mathrm{max}=m_\mathrm{max}=20$ 
and a fourth order Runge-Kutta scheme with step size $\Delta t=0.01$.
For $4\sigma/\epsilon<0.3$, $|\mathsf{P}_1^0|$ and $|\mathsf{Q}_1^0|$ approach 1, 
and we had to increase the number of modes up to $k_\mathrm{max}=m_\mathrm{max}=40$
in order to fully delineate the SNIC line.
The results perfectly match with the location of the codimension-two points, degenerate pitchfork (DP) and
Takens-Bogdanov (TB), which are known analytically.
This implies that the thermodynamic limit of \eqref{eqphasebimodal} is essentially achieved with at most
($2\times2\times40\times41=$) $6560$ degrees of freedom. Actually, ($2\times2\times20\times21=$) $1680$ 
degrees of freedom are generally enough to capture the thermodynamic limit. 
The simulation of the Fourier-Hermite modes does not 
suffer of finite-size fluctuations, in contrast to simulating ensembles of phase oscillators. 
Moreover, the evolution equations are purely algebraic, in contrast to the
computationally expensive trigonometric interaction functions of the oscillator ensemble. 

Another advantage of the moment-based approach is that it 
allows us to track unstable PS states and unstable traveling waves. 
The traveling wave solution corresponds to a solid rotation,
$\mathsf{P}_k^m(t)= \mathsf{p}_k^m e^{ik\Omega t}$, $\mathsf{Q}_k^m(t)= \mathsf{q}_k^m e^{ik\Omega t}$, 
where the angular velocity $\Omega$ is one additional unknown.
After inserting  this solution into Eq.~\eqref{eqmodesbimodal}, the unknowns $\mathsf{p}_k^m$,
$\mathsf{q}_k^m$, and $\Omega$ are found
via a Newton-Raphson algorithm (imposing $\mathsf{p}_1^0\in\mathbb R^+$). 
Moreover, the linear stability of both, PS and traveling waves,
can be accurately determined linearizing the system (for the traveling wave one has to adopt
a rotating reference frame at frequency $\Omega$). 
In the model investigated in this section, 
the previous procedure permitted us to find the locus of a drift-pitchfork bifurcation. 
A part of its locus is indicated by blue dotted line in Fig.~\ref{figbimodal}. 
It is absent in the phase diagram presented in Ref.~\cite{MBS+09}, as there is not an obvious 
manner of observing these unstable solutions simulating an ensemble of oscillators.
At the drift-pitchfork bifurcation, twin unstable traveling waves (born at the Hopf bifurcation) cease to exist,
via collision with an unstable PS state.
An equivalent bifurcation line was recently detected for a distribution 
sum of two Lorentzians \cite{guo20},
although (mistakenly) labeled as a saddle-node bifurcation. 
As the drift-pitchfork line does not involve any attractors we did not made more efforts to
trace it completely. In analogy to the result in
\cite{guo20}, we expected this line to bend backwards and terminate at the origin.
The presence of a drift-pitchfork bifurcation
confirms that the TB point fully consistent with 
the O(2) symmetry of the model: invariance under rotation $\theta\to \theta+c$ and 
reflection $\Omega\to-\Omega$. In particular, the O(2)-symmetric Takens-Bogdanov observed
corresponds to the sceneario ``IV$-$'' in the nomenclature of Ref.~\cite{DN87}, where the 
codimension-2 bifurcation was fully unfolded.

\begin{figure}
	\includegraphics[width=\linewidth]{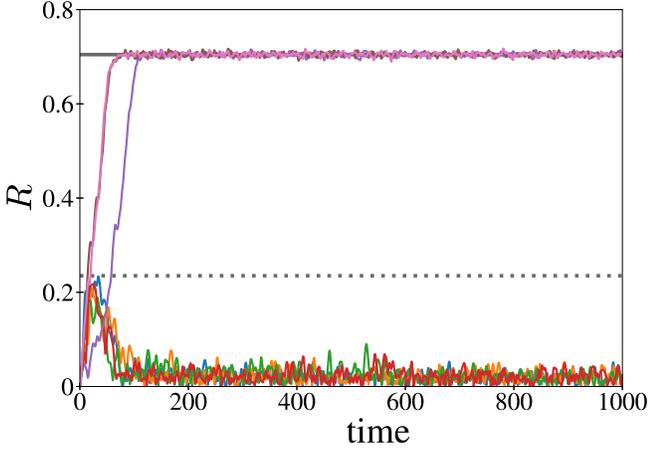}
	\caption{Time evolution of $R(t)$ for $N=5000$ oscillators and 
	7 different initial conditions with random initial phases. The
	parameter values correspond to the cross in Fig.~\ref{figbimodal}(b),
	inside the bistability region UIS/PS, with $\sigma=1$. Stable and unstable PS states
	correspond to fixed points of Eq.~\eqref{eqmodesbimodal}, and they are shown 
	as thick grey lines in solid and dashed styles, respectively.}
	\label{figfluct}
\end{figure}

We conclude this section with a simple numerical experiment.
Our intention is to show the limitations of working 
directly with populations of oscillators. 
We take 5000 oscillators with the parameter values indicated by the cross
in Fig.~\ref{figbimodal}(b). Incoherence is stable, and coexists with PS. Hence,
starting the oscillators in a (finite-size version) of the 
UIS should not be particularly interesting. To our surprise, even 
if the phases are initiated at random, the system may jump to the PS
state. After a surprising initial upstroke of $R$, 3 out of 7 realizations
ended at the PS state, see Fig.~\ref{figfluct}. In this figure, it is manifest the role 
of the unstable PS state, obtained from the moment system, as a threshold 
in the dynamics. 
Needless to say that the moment system \eqref{eqmodesbimodal}
is much more robust describing the thermodynamic limit, due to the lack of fluctuations.
For the same parameter values, incoherence, $\mathsf{P}_k^m=\mathsf{Q}_k^m=\delta_{k,0} \delta_{m,0}$, 
is a stable fixed point for any truncation in $k$ and $m$.

\section{Enlarged Kuramoto model}
In this section, we apply the moment-based approach to a population of phase oscillators with nonpairwise interactions: 
the `enlarged KM'.
This system was recently investigated in Ref.\cite{leon22}. Our purpose here is to explore the $\sigma\to0$ limit of the model, and 
the convergence of the Lyapunov exponent as $N\to\infty$ in a regime of collective chaos.
We start writing the coupling function:
\begin{multline}\label{eq2ndordermodel}
 G(\theta,t)=
 \epsilon \eta \,  R  \, \sin(\Psi-\theta_j+\alpha) 
 + \frac{\epsilon^2 \eta^2}{4} 
 \bigg[R\sin(\Psi-\theta_j+\beta)\\
 - R^2 \sin(2\Psi-2\theta_j+\beta)+   R \, Q \sin(\Phi-\Psi-\theta_j) 
 \bigg] .
\end{multline}
Two mean fields enter in this equation
$Z_1\equiv R \, e^{i\Psi}$, and the
second Kuramoto-Daido order parameter $Z_2\equiv Q \, e^{i\Phi}$.
In contrast to other models with nonpairwise interactions, see e.g.~\cite{tanaka11,KP15,SA19}, the enlarged KM
is not simply postulated.
It is obtained applying phase reduction\cite{leon19} to a population of Stuart-Landau oscillators, 
up to second order in the coupling constant $\epsilon$. 
Constants $\eta$, $\alpha$ and $\beta$ in Eq.~\eqref{eq2ndordermodel}
depend on the original constants $c_1$ and $c_2$  via
$\eta\equiv\sqrt{(1+c_2^2)(1+c_1^2)}$, 
$\alpha\equiv\arg[1+c_1 c_2+(c_1-c_2)i]$, and
$\beta\equiv\arg(1-c_1^2+2 c_1 i)$. 
For completeness, we write the ODE governing the Stuart-Landau oscillators. In this way, 
we can understand the meaning of $c_1$ and $c_2$ (related to reactivity and shear, respectively):
 \begin{equation}\label{CGLE}
 \dot{A}_j= (1+i\sigma\omega_j)A_j - ( 1 + i c_2) |A_j|^2 A_j+\epsilon(1+ic_1) \left(\overline{A}- A_j\right),
 \end{equation}
 where $\overline{A}=N^{-1}\sum_j A_j$. 

In a previous work\cite{leon22}, we showed that Eq.~\eqref{eq2ndordermodel} reproduces the 
rich phenomenology of the ensemble of Stuart-Landau oscillators at weak coupling. In contrast, 
the first-order phase approximation, neglecting powers of $\epsilon$ above or equal to 2, 
only predicts two different collective states: UIS and PS.
The enlarged KM  cannot be analyzed within the Ott-Antonsen theory,
because of the second harmonic in $\theta$ in the interaction function.
This means that no low-dimensional description is available, 
irrespective of the frequency distribution. As we will show below, under weak coupling and weak heterogeneity 
($\sigma\ll1$, $0<\epsilon\ll1$)
the study of the model through direct numerical simulations is impractical. Apart from the
unavoidable finite-size fluctuations,
the dynamics turns out to be very slow, and long transients are needed to reach stationary regimes. 
Possessing an efficient numerical scheme becomes essential.
We resort to the Fourier-Hermite modes to explore the dynamics of the enlarged KM with Gaussian heterogeneity.
After, straightforward calculations we obtain the evolution equations of the moments $\mathsf{P}_k^m$:
	\begin{eqnarray}
	{k}^{-1}\dot{\mathsf{P}}_k^m&=& \sigma\left(\sqrt{m} \mathsf{P}_k^{m-1}-\sqrt{m+1}\mathsf{P}_k^{m+1}\right) \nonumber\\
	&+&\frac{\epsilon \eta}{2}\left(\mathsf{P}_{k-1}^mZ_1e^{i\alpha}-\mathsf{P}_{k+1}^m Z_1^* e^{-i\alpha} \right) \nonumber\\
	&+&\frac{\epsilon^2\eta^2}{8}\left(
	\mathsf{P}_{k-1}^m Z_1 e^{i\beta}-\mathsf{P}_{k+1}^m Z_1^* e^{-i\beta}-
	\mathsf{P}_{k-2}^m Z_1^2 e^{i\beta} \right.\nonumber \\ 
	&+& \left.\mathsf{P}_{k+2}^m Z_1^{*2} e^{-i\beta}+
	\mathsf{P}_{k-1}^m Z_2 Z_1^* -\mathsf{P}_{k+1}^m Z_2^* Z_1 \right),
	\label{odes}
	\end{eqnarray}
where $Z_1=\mathsf{P}_{1}^0$ and $Z_2=\mathsf{P}_{2}^0$.
These equations were integrated using the fourth order Runge-Kutta method with time step
$\Delta t=0.01$.
	
\begin{figure}
	\includegraphics[width=\linewidth]{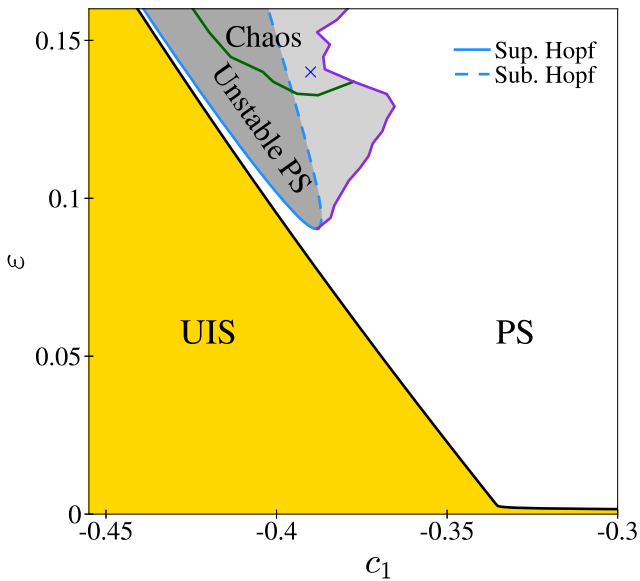}
	\caption{Partial phase diagram obtained for the interaction function	
	\eqref{eq2ndordermodel}, with for $c_2=3$ and $\sigma=10^{-3}$. Yellow and white regions 
	indicate stable UIS and stable PS, respectively. 
	At the solid (dashed) blue line PS undergoes a supercritical (subcritical) Hopf bifurcation. 
	In the light grey shaded region PS and more complex (unsteady) dynamics coexist. 
	The green line bounds the chaotic domain inside the unsteady-dynamics region.
	}
	\label{figdiag}
\end{figure}
	
\subsection{Phase diagram}
We start reviewing the phase diagram obtained in Ref.~\cite{leon22}.
Figure~\ref{figdiag} shows a partial phase diagram for specific values of the parameters $c_2=3$ and $\sigma=10^{-3}$. 
A couple of bifurcation lines are omitted to simplify the presentation.
In the yellow region UIS is stable. Crossing the black line 
stability is transferred to a partially synchronized state (PS). 
Interestingly, the blue line indicates the locus of a secondary instability where PS itself
becomes unstable through a Hopf bifurcation. The linear stability of PS was calculated 
finding the rotating solution $\mathsf{P}_k^m(t)= \mathsf{p}_k^m e^{ik\Omega t}$, 
as with the traveling wave in the previous section, and evaluating the
eigenvalue of the Jacobian with the largest real part in the corotating frame.
Hence, PS is unstable inside the dark shaded region of Fig.~\ref{figdiag}.
This instability is only possible thanks to the last two terms in Eq.~\eqref{eq2ndordermodel}, which 
confer three-body phase interactions to the model. Hence, in the shaded region, 
the meanfield dynamics display complex oscillations. 
In particular, above the green line the dynamics becomes chaotic
(or hyperchaotic), see \cite{leon22}. The chaoticity is characterized 
calculating the Lyapunov exponents of the dynamical system \eqref{odes}.
We end this overview noting
that in the light shaded region there is bistability between PS and unsteady dynamics (chaotic or not). 

\begin{figure}
	\includegraphics[width=\linewidth]{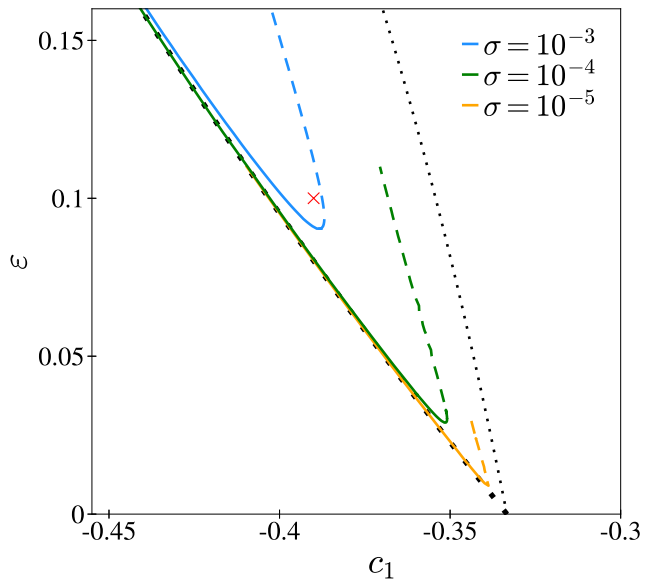}
	\caption{Location of Hopf bifurcation for $c_2=3$ and $\sigma=10^{-3},10^{-4},10^{-5}$ in blue, green and orange respectively. The lines were obtained using \eqref{odes} with $k_{\mathrm{max}}=m_{\mathrm{max}}=40$. The right boundary is not completely depicted 
	because an extraordinary large number of modes would be needed to calculate it. 
	The black dotted lines are the stability boundaries of UIS (left) and full synchrony (right) for $\sigma=0$.
	The red cross indicates the parameter values used in Fig.~\ref{fig_sf}.
	}
	\label{figsigmas}
\end{figure}

It is important to stress that obtaining Fig.~\ref{figdiag} was possible 
thanks to the use of the Fourier-Hermite modes $\mathsf{P}_k^m$,
with a maximal size of  $k_\mathrm{max}=m_\mathrm{max}=40$.
Performing direct numerical simulations it would be virtually impossible
to achieve such a level of detail in the phase diagram. In particular,
to validate the existence of collective chaos is particularly difficult, see below.

\subsection{The homogeneous  limit ($\sigma\to0$)}

The case $\sigma=0$ 
was studied in \cite{leon19}. That system only exhibits UIS, 
full synchrony, and an intermediate region with nonuniform
incoherent states ($Z_1=0$, $|Z_2|\equiv Q=\mathrm{const.}>0$). There is not a correspondence between
$\sigma=0$ and $\sigma>0$. While, full synchrony is the limit of PS, it is not obvious how
nonuniform incoherent states may appear as $\sigma$ is lowered to 0.
Next, we use the Fourier-Hermite modes to understand the limit $\sigma\rightarrow0$.

\begin{figure}
	\includegraphics[width=\linewidth]{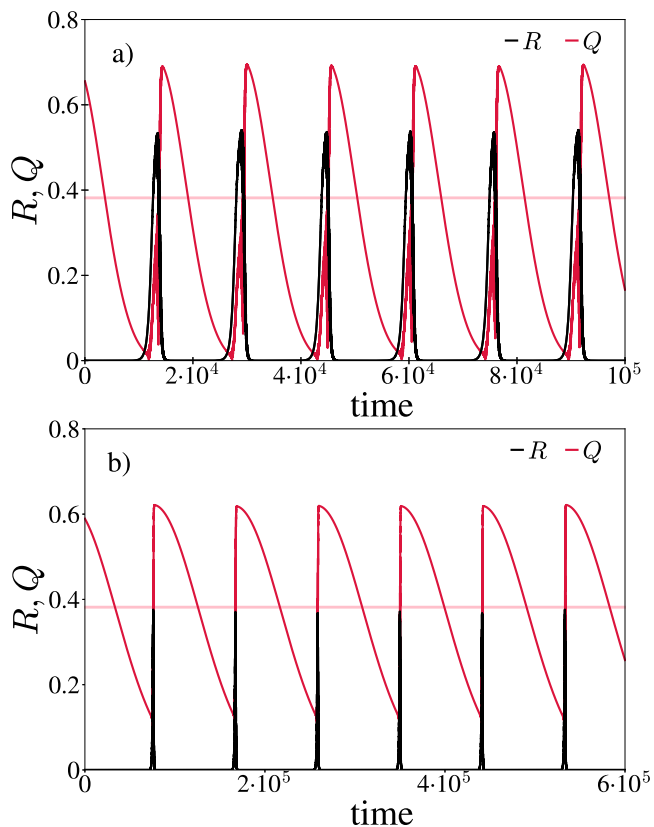}
	\caption{Dynamics of the enlarged KM for (a) $\sigma=10^{-4}$ and (b) $\sigma=10^{-5}$.
	The time series are the result of integrating Eq.~\eqref{odes} with $c_1=-0.39$ and $\epsilon=0.1$ (see the red cross
	in Fig.~\ref{figsigmas}). The horizontal pink line is the value of $Q$ in the limit $\sigma=0$, see Eq.~(27)
	in Ref.~\cite{leon19}. Note the different time scale of both panels.
	}
	\label{fig_sf}
\end{figure}

In Fig.~\ref{figsigmas} we show the stability boundary of PS for three different $\sigma$ values. 
Let us emphasize that finding those three boundaries from direct numerical simulations of an ensemble of
oscillators would be virtually impossible, specially considering small values of $\sigma$ and $\epsilon$
achieved.
We may see that as $\sigma$ decreases, the tip of the boundary progressively approaches the abscissa at $c_1=-1/3$. 
We conclude that as
soon as $\sigma$ becomes nonzero the Hopf bifurcation appears making the $\sigma\to0$ limit singular.
The left (right) black dotted line in the figure is the stability boundary of UIS (full synchrony)
for $\sigma=0$. 
From the figure we infer that the supercritical Hopf bifurcation of PS 
collides with UIS boundary in the limit $\sigma\rightarrow0$. Complementary, 
the subcritical branch becomes the stability boundary of full synchrony. 
This is consistent because bistability is also observed in the 
$\sigma=0$ case \cite{leon19}.

The previous discussion, at the level of bifurcations, is not saying us anything 
about how is the system behaving as $\sigma\to0$. In Figs.~\ref{fig_sf}(a)
and \ref{fig_sf}(b) the time evolution of $R(t)$ and $Q(t)$ are depicted
for $\sigma=10^{-4}$ and $10^{-5}$, respectively. The remaining
parameter values are the same in both panels, and correspond to the 
red cross included in Fig.~7.
The model displays slow-fast dynamics.
The mean field $R$ remains close to 0 most of the time, while $Q$
exhibits a slow decay, followed by a rapid increase. The different
time scale in both panels indicates that the slow time scale
is diverging as $\sigma\to0$. Moreover the range of $Q$ decreases,
and (slowly) approaches the asymptotic value for $\sigma=0$ \cite{leon19}, indicated by a pink horizontal line.

\subsection{Collective chaos}

In sharp contrast to the standard KM and other models of phase oscillators, the enlarged KM ---defined by Eqs.~\eqref{eqphaseosc} and \eqref{eq2ndordermodel}--- 
exhibits collective chaos with a unimodal distribution of the natural frequencies.
Figure~\ref{figdiag} shows
the location of the chaotic region for particular values of $\sigma=10^{-3}$ and $c_2=3$.  
In order to determine the boundary of chaos in Fig.~\ref{figdiag}, we needed to determine
the largest Lyapunov exponent in the thermodynamic limit. Again, integrating a finite number of phase oscillators is quite unproductive, 
due to the ubiquitous microscopic phase chaos \cite{popovych05}, 
which yields a macroscopic 
amount, i.e.~$O(N)$,
of positive Lyapunov exponents (not shown). 
In contrast, the Fourier-Hermite moments do not suffer of microscopic phase chaos. It
is therefore possible to fully characterize the collective chaos of the enlarged KM, 
and to determine its boundary.

We present next, a numerical test supporting our previous assertions
for a specific set of parameters: $c_2=3$, $c_1=-0.39$, $\epsilon=0.14$, and $\sigma=10^{-3}$,
 see the blue cross in Fig.~\ref{figdiag}. 
For these parameter values two stable states coexist in the thermodynamic limit: collective chaos and PS.
In the chaotic state the Lyapunov spectrum, obtained linearizing \eqref{odes} and applying 
Bennetin's algorithm, is $\{\lambda_n\}_{n=1,2\ldots}= \{1.26\times 10^{-4},6.31\times 10^{-5},1.38\times 10^{-5},0,\ldots \}$. This spectrum contains three positive Lyapunov exponents, 
indicating collective
hyperchaos. This means that, for a sufficiently large ensemble, the positive part of the 
Lyapunov spectrum
will consist of three exponents neatly above zero, supplemented by a quasi-continuous set 
of exponents of $O(N)$ size (whose values approach zero as $N\to\infty$).

\begin{figure}
	\includegraphics[width=\linewidth]{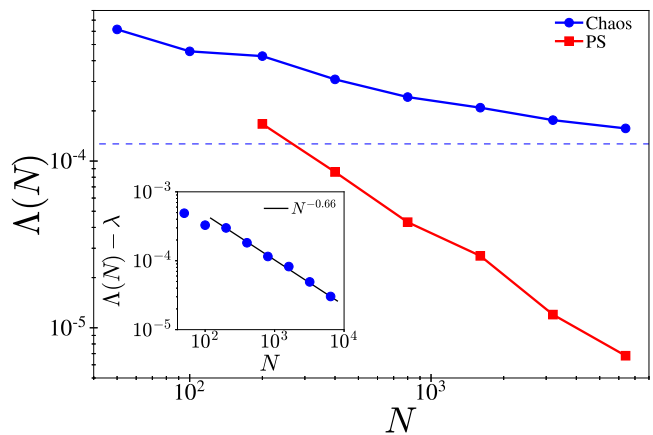}
	\caption{Largest Lyapunov exponent $\Lambda$ as function of the population size $N$
	for the enlarged KM and two different coexisting states. The values of $\{\omega_j\}_{j=1,\ldots,N}$ were selected deterministically to represent the Gaussian distribution \eqref{gauss}.
	The parameter values are those marked by a blue cross in Fig.~\ref{figdiag}. Circles and squares correspond, respectively, 
	to the states of collective chaos and PS in the thermodynamic limit. The value of the largest Lyapunov exponent
	$\lambda_1=\lambda$, obtained from the moment system \eqref{odes} in the chaotic state, is indicated by a horizontal dashed line. 
	The inset shows a log-log plot of the distance to the
	asymptotic Lyapunov exponent $\lambda$  vs.~the system size $N$. The linear fitting, 
	depicted by a solid line,
	was performed considering points with $N\ge200$. Note that $\lambda$ is not a fitting parameter.
	}
	\label{figlyap}
\end{figure}

Now, we turn our view to the results obtained from 
direct numerical simulations with $N$ oscillators. In Fig.~\ref{figlyap}
we represent the largest Lyapunov exponent $\Lambda$ as a function of $N$ for each
of the coexisting states (PS and collective chaos).  
In both cases the Lyapunov exponent decreases as $N$ grows. For PS we find a clean
decay to zero as $\Lambda(N)\sim1/N$. This behavior is known for the UIS
in the KM \cite{popovych05} (logarithmic corrections may be present depending on the
sampling procedure of the natural frequencies\cite{carlu18}),
and it can be arguably expected for PS too.
In the state of collective chaos, the 
decrease of $\Lambda(N)$ appears to saturate at a finite value consistent with 
$\lambda\equiv\lambda_1=1.26\times10^{-4}$, obtained 
from the Fourier-Hermite modes. To further confirm this guess,
we represent in log-scale $\Lambda(N)-\lambda$ vs.~$N$ in the inset of
Fig.~\ref{figlyap}. The data are fully consistent with a power-law 
convergence $\Lambda(N)-\lambda\propto N^{-\gamma}$.
The exponent $\gamma$, estimated by a linear fit, 
turns out to be nontrivial: $\gamma\simeq0.66$. Further work is needed
to assess whether and how the value of $\gamma$ depends on parameters.

\section{Conclusions and Outlooks}

In this work, we have 
studied a
moment-based approach for ensembles of globally coupled 
phase oscillators with Gaussian distributed natural frequencies. 
Several truncations and polynomial closures of the moment system 
have been tested in the KM. 
For the index $m$ (related to the decomposition of the oscillator 
density in the frequency variable), the linear closure 
is readily implemented, and 
apparently reliable in all analyzed cases.

The moment-based approach
allows us to describe
the collective dynamics of an infinite population 
with a relatively small number of ODEs,
avoiding in this way finite-size flucutations, inherent to direct numerical simulations. 
Within this framework 
numerical continuation of PS states is possible, irrespective of their stability.
Moreover, linear stability analysis and the computation of  
``collective'' Lyapunov exponents 
become simple tasks.


We have applied the moment-based approach, specifically Fourier-Hermite modes, 
to two problems. 
The first one, the KM with a frequency distribution given by the sum of two Gaussians,
previously considered in \cite{MBS+09}. Here, using 
moment-based dynamics we have refined the phase diagram in \cite{MBS+09},
obtaining the accurate loci of bifurcations. Moreover, a new bifurcation
line (drift-pitchfork) has been detected, completing the picture
around the Takens-Bogdanov point.

The second problem addressed in this work is a complicated phase model,
the enlarged KM,
derived via second-order phase reduction from an ensemble of Stuart-Landau oscillators.
For this system, we have focused on asymptotic properties 
hardly discernible with direct simulations of large ensembles. 
First, we have examined the
change of the stability boundary of PS as the heterogeneity becomes
less and less pronounced ($\sigma\to0$). We found
the convergence to two stability boundaries analytically 
known for $\sigma=0$ \cite{leon19}. The limit $\sigma\to0$ is 
singular, yet consistent.
For small $\sigma$  the system displays slow-fast dynamics, 
with the duration of the slow phase diverging as $\sigma\to0$.
In second place, we investigated the state of collective chaos and
the (power-law) convergence of the largest 
Lyapunov exponent $\Lambda(N)$ with the size. The knowledge of the 
asymptotic value $\lambda$ from the moment-based simulation
was crucial to estimate the exponent $\gamma\simeq0.66$ of the power-law behavior:
$\Lambda(N)-\lambda\sim N^{-\gamma}$.

Our presentation is limited to the simplest situations.
We provide next a list of manners in which 
the moment-based approach can be extended:
\begin{enumerate}
 \item We have restricted to Gaussian heterogeneity. Other distributions $\tilde g(\omega)$
with finite moments can be analyzed using their corresponding
sets of orthonormal polynomials $\{{\tilde h}_m(\omega)\}_{m=0,1,\ldots}$, satisfying
$\int {\tilde h}_m(\omega) {\tilde h}_n(\omega) {\tilde g}(\omega) d\omega=\delta_{m,n}$. 
Each set of orthonormal polynomials satisfies a specific recurrence relation:
$\omega {\tilde h}_m(\omega)= b_m {\tilde h}_{m+1}+a_m {\tilde h}_m+b_{m-1}{\tilde h}_{m-1}$,
leading to a particular variation of Eq.~\eqref{eqdotmodes}. (Note that $a_m=0$ if ${\tilde g}(\omega)$
is even.) Still, for the specific $\tilde g(\omega)$, a preliminary study of the suitable closure(s) should be carried out.
\item We have considered purely deterministic equations.
Adding independent white noises $+\xi_j$ to Eq.~\eqref{eqphaseosc} 
does not modify the approach essentially. If the  
covariance of the noise is $\langle\xi_j(t) \xi_{j'}(t')= 2 D \delta_{j,j'} \delta(t-t')$,
then the continuity Eq.~\eqref{cont} 
gains the term  $+D\partial^2 \rho/\partial\theta^2$ in the
right-hand side (Fokker-Planck equation). This simply 
results in
a new (dissipative) 
term $-D k\mathsf{P}_k^m$ in Eq.~\eqref{eqdotmodes}. 
\item In our presentation heterogeneity appears in an additive form, see Eq.~\eqref{eqphaseosc}.
The same strategy could be followed, in principle, if multiplicative
disorder was present instead. Models of this type include ensembles of theta neurons \cite{LBS13,LBS14}, generalizations of
the KM \cite{HS11,MP11,PM11,MP11p,IPM+13,IMS14}, and the Winfree model with distributed phase response curves \cite{PMG19}.
\end{enumerate}

All in all, we judge the moment-based approach studied in this paper as a
very useful tool to investigate the dynamics of ensembles 
of heterogeneous phase oscillators.

 \begin{acknowledgments}
IL acknowledges support by Universidad de Cantabria and 
Government of Cantabria under the Concepci\'on Arenal programme.
 \end{acknowledgments}

\section*{DATA AVAILABILITY}
 
 The data that support the findings of this study are available from the corresponding author 
 upon reasonable request.
 

\section*{References}

%

\end{document}